%% file: eventlines-arxiv.tex
\documentclass[11pt,twocolumn]{article}

\usepackage{times}

\usepackage[labelfont=bf]{caption}

\usepackage[top=0.5in,bottom=0.75in,left=0.5in,right=0.5in]{geometry}

\usepackage{url}

\usepackage{subfig}

\usepackage{graphicx}

\usepackage{amsmath}

\graphicspath{{figures/}{plots/}{.}}

\usepackage[numbers]{natbib}
\usepackage[bookmarksopen,bookmarksnumbered,colorlinks=true,allcolors=blue]{hyperref}

\usepackage{xcolor}

\usepackage[compact]{titlesec}

\usepackage{helvet}
\usepackage{sectsty}
\allsectionsfont{\sffamily}

\usepackage{mathptmx}   

\usepackage{flushend}

\usepackage{framed}

\usepackage[strict]{changepage}

\usepackage{subfig}

\begin{document}


\title{\sffamily EventLines: Time Compression for Discrete Event Timelines}

\author{Yuetling Wong$^{1}$  and Niklas Elmqvist$^{2}$\\
\scriptsize $^1$Purdue University, West Lafayette, IN, USA;
\scriptsize $^2$Aarhus University, Aarhus, Denmark}


\maketitle

\begin{abstract}
    \input{contents/abstract}
\end{abstract}



\input{contents/01-introduction}
\input{contents/02-background}
\input{contents/03-technique}
\input{contents/04-studies}
\input{contents/05-results}

\input{contents/06-discussion}
\input{contents/07-application}
\input{contents/08-conclusion}

\section*{Acknowledgments}

This work was partially supported by the U.S.\ National Science Foundation grant TUES-1123108.
Any opinions, findings, and conclusions, or recommendations expressed here are those of the authors and do not necessarily reflect the views of the funding agencies.


\end{document}

%% file: contents/abstract.tex
Discrete event sequences serve as models for numerous real-world datasets, including publications over time, project milestones, and medication dosing during patient treatments.
These event sequences typically exhibit bursty behavior, where events cluster together in rapid succession, interspersed with periods of inactivity.
Standard timeline charts with linear time axes fail to adequately represent such data, resulting in cluttered regions during event bursts while leaving other areas unutilized.
We introduce EventLines, a novel technique that dynamically adjusts the time scale to match the underlying event distribution, enabling more efficient use of screen space.
To address the challenges of non-linear time scaling, EventLines employs the time axis's visual representation itself to communicate the varying scale.
We present findings from a crowdsourced graphical perception study that examines how different time scale representations influence temporal perception.

%% file: contents/01-introduction.tex

\section{Introduction}

Minard's seminal 1869 map visualizes Napoleon's catastrophic Russian campaign of 1812, masterfully integrating temporal and spatial dimensions~\cite{Tufte1992}.
The LifeLines system~\cite{PlaisantMilashRoseWidoffShneiderman1996} offers a comprehensive visualization framework for personal histories through individual timelines.
ThemeRiver~\cite{Havre2000} employs a ``river'' metaphor flowing through time to illustrate thematic variations across document collections.
While these visualizations demonstrate the power of timeline-based temporal data representation, linear timelines become problematic when handling irregularly sampled or bursty data.
Consider an emergency room scenario: some days see minimal patient flow, while others experience overwhelming surges due to accidents, seasonal outbreaks, or emergencies.
Visualizing such \textit{bursty event data} through traditional timelines, with their uniform temporal scaling, creates an inefficient contrast of densely packed and sparse intervals, proving particularly challenging for large datasets.

To overcome these limitations, we introduce \textit{EventLines}, a novel timeline visualization optimized for bursty event data that maximizes temporal axis display space utilization.
Rather than employing linear temporal scaling, we divide the time axis into piece-wise linear segments between adjacent events (with simultaneous events grouped as single units).
These segments receive equal display space on the timeline.
This approach ensures maximum separation between temporally proximate events, thereby reducing visual clutter and enhancing readability for bursty event data.

\begin{figure*}[htb]
  \includegraphics[width=\linewidth]{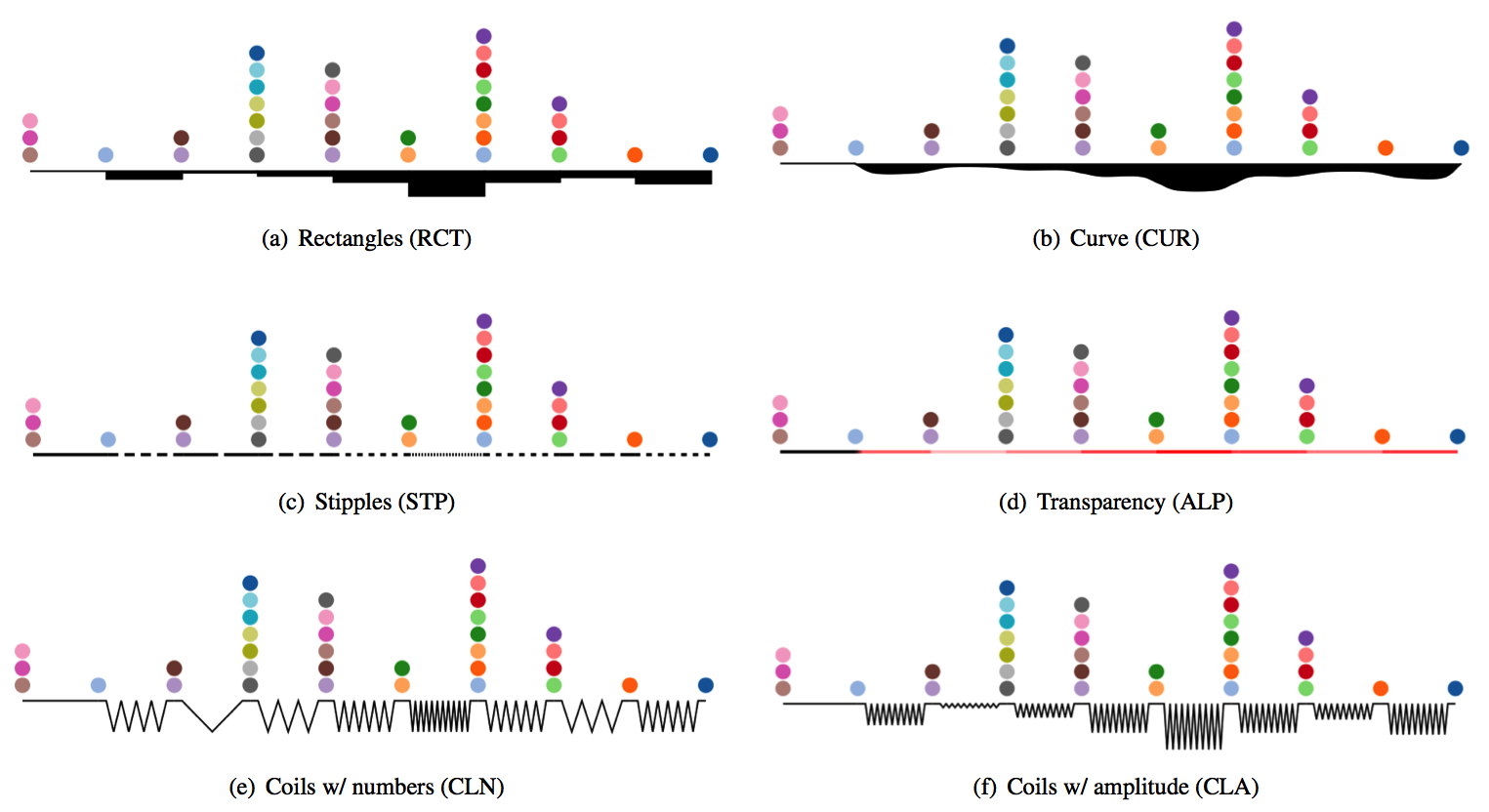}
  \caption{\textbf{EventLines visual representations.}
  These examples all use data-aware, piece-wise linear compression of the time axis to improve the use of available display space.
  Each example shows a different visual representation used for conveying the compression of the time axis.}
  \label{fig:teaser}
\end{figure*}

However, varying scaling across the timeline poses challenges for estimating specific times and durations.
To address this, the EventLines technique incorporates visual cues within the timeline itself to convey non-linear temporal scaling, enabling viewers to intuitively grasp the non-linear nature through visual inspection alone.
We designed six candidate visualizations to represent the time scaling (Figure~\ref{fig:teaser}): using the axis thickness as curves (CUR) or rectangles (RCT); using transparency (ALP); using coils on the axis with varying number (CLN) and amplitude (CLA); and using stipples (STP).
Figure~\ref{fig:teaser} shows examples of each representation.
We implemented all of these techniques in an EventLines plugin for the D3 toolkit~\cite{Bostock2011}.

To evaluate these representations, we conducted a crowdsourced graphical perception study~\cite{HeerBostock2010} comparing their performance with typical bursty event data.
Our results indicate that three techniques---number coils (CLN), stipples (STP), and rectangles (RCT)---demonstrate superior performance across representative timeline tasks, including absolute time estimation, interval estimation, and time comparison.
As expected, despite our visual scaling representations, participants indicated a preference for traditional timelines when performing absolute time estimation tasks.

%% file: contents/02-background.tex

\section{Related Work}

Stock market prices, patient drug experiment reports, even schedules of our everyday activities are all time-series data.
Accordingly, myriad visualization techniques have been designed to address different aspects of such data.
In this section we review relevant literature on visualizing such data.

\subsection{Discrete Event Timelines}

Discrete event data often have interesting relationships or uneven distribution within or across time.
In addition, the amount of data may be large and each data point can have multiple discrete properties.
To visualize event data, we cannot just connect events to form curves as for continuous data.
Many visualizations have been designed to address these problems.

LifeLines~\cite{PlaisantMilashRoseWidoffShneiderman1996} is a general
visualization for personal histories.  It provides a zoomable
single-screen overview to reduce the chance of missing information; a
timeline visualization with color encoding, silhouettes, and shadows,
as well as hierarchy labels to facilitate the spotting of anomalies
and trends and streamline the access to details; and a simple and
tailorable interface to lower the learnability barrier.  Its follow-up
project LifeLines 2~\cite{Wang2009} visualizes temporal categorical
data across multiple records for pattern exploration and discovery to
support hypothesis generation, and finding cause-and-effect
relationships using three operators: align, rank, and filter.  Two
related projects are LifeFlow~\cite{Wongsuphasawat2011}, which
provides a novel interactive visual overview of event sequence for any
number of records, and EventFlow~\cite{Monroe2014}, which displays
point and interval events using a comprehensive and innovative
visualization.

Another interactive visualization dealing with patient data is
CareVis~\cite{AignerMiksch2006}, which provides multiple simultaneous
views to cover different aspects of the complex underlying data
structure. It contains three views: (1) a logical view showing flow
charts; (2) a temporal view based on the
LifeLines~\cite{PlaisantMilashRoseWidoffShneiderman1996} concept and a
zoomable timeview~\cite{AignerMikschThumberBiffl2005} to display
hierarchical decomposition of temporal intervals and complex time
annotations; and (3) a so-called QuickView panel used to monitor
important values at a prominent positions.  A similar multiview
timeline visualization is Continuum~\cite{Andre2007}, which provides
an overview panel displaying the core data as a histogram to scale up
to large datasets, a detail view panel showing hierarchy information
of data selected by the viewfinder of overview panel, and a dimension
filter panel controlling the level and type of data displayed.  One
feature that makes Continuum remarkable is the arching connection
lines that connect relative information between two split detail
panels.

\subsection{Distortion and Compression}

The large volume of time-series data sometimes makes a timeline
visualization extremely long and dense.  To allow users to
focus on specific data points while maintaining the overview for the
whole data, several distortion and compression techniques have been
presented in the literature.

Focus+context techniques~\cite{Furnas1986} represent one family of a
particularly powerful approach that distorts the visual space so that
uninteresting parts of the visualization are compressed to provide
more space for the focus area.  Such distortion and compression
techniques can be divided into two categories: geometric distortion
and semantic distortion~\cite{ElmqvistHenryRicheFekete2008}.  Examples
of geometric distortion include Table Lens~\cite{Rao1994}, which
applies fisheye distortions to a tabular visualization, and Document
Lens~\cite{Robertson1993}, which visualizes a large document aligned
as a rectangle array of pages with a 3D focus region.
TreeJuxtaPozer~\cite{Munzner2003} is a visualization system designed
for large tree comparison using an extension of the rubber sheet
stretching metaphor known as accordion drawing.  Other applications
involving space distortion include
M\'{e}lange~\cite{ElmqvistHenryRicheFekete2008}, which folds the
intervening space to guarantee visibility of multiple focus regions,
and stack zooming~\cite{JavedElmqvist2010}, which combines the
techniques of focus+context, split-screen, overview+detail, and
hierarchical navigation to support multi-focus interactions while
retaining context and distance awareness.  Finally,
DOITree~\cite{Card2002} and
SpaceTree~\cite{PlaisantGrosjeanBederson2002} are both examples of
applying semantic distortion to hierarchical data.

Compared to all of these approaches, our new EventLines time
distortion technique is data-driven in that it automatically adapts
the distortion based on the underlying event data.  In this way, our
work is more similar to topology-aware navigation techniques for
graphs, such as the techniques proposed by Moscovich et al.~\cite{Moscovich2009}

\subsection{Graphical Perception}

The concept of \textit{graphical perception} is often used to examine
the effectiveness of different timeline visualizations.  Graphical
perception is defined as users' ability to comprehend the visual
encoding and thereby decode the information presented in the
graph~\cite{Lohse1991}.~\cite{Cleveland1984} identified a set of 
elementary perceptual tasks for people to extract quantitative information and 
an ordering of accuracy of these tasks.~\cite{HeerKongAgrawala2009} performed two
controlled experiments to compare line charts with horizon graphs,
measuring the speed and accuracy of subjects' estimates of value
differences between charts.  \cite{JavedMcDonnelElmqvist2010} went
beyond this work, exploring the performance of multiple time series
visualization with three tasks: comparing the maximum, finding the
highest slope, and determining the highest value at a specific point
of each series.

Crowdsourcing presents an attractive option for evaluating graphical
visualization with its low cost, scalability, and population
diversity.  However, validation is required before researchers can use
this option to conduct user studies.  \cite{Kosara2010} conducted
several conception and cognition studies on Amazon's Mechanical Turk,
where the study of examining users' understanding of two tree
visualization methods was identical to a previous study they performed
in the lab.  Based on the experiment process and study results, they
validate the reliability of data by using qualification to filter
eligible turkers and present ways to avoid common problems by taking
them into account in the study design.  Similarly,
\cite{HeerBostock2010} performed several previous and new experiments
on Amazon's Mechanical Turk and validated the viability, accuracy, and
budget of using crowdsourcing to conduct traditional graphical
perception studies.

%% file: contents/03-technique.tex

\section{Time Distortion using EventLines}

We propose \textit{EventLines}, a novel visualization technique to leverage the high density of bursty event data by making the optimal use of available display space through distortion, as well as conveying time compression to the viewer using the time axis.
Our work is influenced by that of Shannon et al.~\cite{Shannon2009}, but goes beyond dynamic data flow and systematically studies visual representations for this data.
In the following subsections, we first discuss the challenges of bursty data.
This is followed by description of our time compression technique, compression visualization, and details about our implementation.

\subsection{Challenges of Bursty Data}

\textit{Bursty data} is defined as temporal data with significant clustering, i.e., where multiple events occur in rapid succession during short time intervals, separated by periods of relative inactivity.
Common examples include social media posts during major events, network traffic during peak hours, or patient arrivals in emergency rooms during disasters.

Traditional timeline visualizations struggle with bursty data because they allocate screen space proportionally to time duration.
This linear time mapping creates two significant problems in areas of event bursts:
First, events become densely packed and often overlap, making individual elements indistinguishable.
Second, the limited space between events makes interactive elements, such as tooltips or time labels, difficult to access and read.
Conversely, periods of inactivity consume valuable screen space while conveying little information.
This inefficient space utilization becomes particularly problematic when visualizing datasets that span long time periods or contain frequent bursts.
Furthermore, the stark contrast between densely packed and sparse regions can make it challenging for users to:
(1) identify temporal patterns within bursts,
(2) compare the temporal distribution of events across different bursts, and
(3) understand the relationship between events during transition periods between busy and quiet intervals.



\subsection{Solution: Time Compression}

To address the clutter problem of bursty data, instead of using the
traditional linear timeline to align the data points, we divide the
timeline into equal intervals so that any two groups of data with
adjacent values of $t_i$ have the same space between them.  More
specifically, given a dataset with n different values of $t_1$ \ldots
$t_n$, in a traditional linear timeline, the length of interval
between $t_i$ and $t_{i+1}$ is proportional to the total length of the
timeline which equals \textit{total\_length}
$*~(t_{i+1}-t_i)/(t_n-t_1)$.  For our visualization, the length of
each interval will be the same---$(t_n-t_1)/(n-1)$---which is no
longer linear for the whole time axis.  However, within each interval,
the scale is linear and the same as the minimum of the n-1 intervals.

An example with 40 data points is shown in the lower part of
Figure~\ref{fig:solution}.  We can see from the figure that bursty and
sparse data in the visualization are distributed evenly.  This
improves the readability of individual data points and makes it easier
to distinguish particular events.  However, the downside of this
method is that the straightforward linear scaling across the entire
timeline is now lost: without looking at the labels on the timeline
axis, users have no idea about how much time each interval represents
or which interval represents the longest amount of time.  This problem
is caused by the non-linearity of the time axis, which means losing
the common distance information between two time events.

\subsection{Solution: Temporal Axis Visualization}

To convey the piece-wise linear scaling of the temporal axis, we
propose to use the visual representation of the time axis itself.
More specifically, our goal is to go beyond the traditional
representation of the time axis as a straight line with tick marks,
and instead use it to convey the changing scale along the axis.  The
visual representation can rely on each period being piecewise linear,
i.e.\ the scaling is constant in the entire period.

We derived six visual representations for the time axis for this
purpose (Figure~\ref{fig:teaser}):

\begin{itemize}

\item\textbf{Axis thickness}: The thickness of the line representing
  the axis can convey the amount of compression: a thicker line means
  higher compression, and a thin line means none.  We devise two ways
  to transition between adjacent regions:
  
  \begin{itemize}
  \item\textbf{Rectangles (RCT)}: The transition from one period to
    another is a step function, resulting in a rectangular shape for
    the time axis (Figure~\ref{fig:teaser}a).
    
  \item\textbf{Curves (CUR)}: The transition is a sine function,
    resulting in a curved shape from one period to the next
    (Figure~\ref{fig:teaser}b).
  \end{itemize}	

\item\textbf{Axis appearance}: The visual appearance of the time axis
  can also be modified to convey additional information. We identify
  two specific approaches that are viable for this:

  \begin{itemize}
  \item\textbf{Stipples (STP)}: Here we use one-dimensional stippling
    of the line representing the axis to convey the amount of time
    that has been compressed.  A solid line means no compression (in
    order for the technique to be consistent with a simple linear time
    axis), whereas an increasing number of increasingly smaller dashes
    means increased compression (Figure~\ref{fig:teaser}c).
    
  \item\textbf{Transparency (ALP)}: The alpha value for the axis line
    can also be used to convey compression; again, for consistency, we
    use an opaque black line to convey no compression, and a red line
    whereas decreasing amounts of transparency (linearly increasing
    alpha) will convey a higher degree of compression.  This makes
    sense because it is intuitive that more time units squeezed into
    the same space would make the line ``heavier''.  So, we choose
    this mapping to maintain consistency with traditional simple
    timelines (Figure~\ref{fig:teaser}d).
  \end{itemize}	
	
\item\textbf{Glyphs}: In statistical graphics, the convention for
  graphical axes that do not start at zero is to a draw a coil-like
  glyph on the axis to communicate that the unit has been compressed.
  We adopt the same idea here to timeline compression, deriving two
  different ways to achieve this:

  \begin{itemize}
  \item\textbf{Coils w/ numbers (CLN)}: The number of distinct coils
    in a period communicates the amount of time that has been
    compressed; a higher number of coils means a larger amount of time
    compression is in effect (Figure~\ref{fig:teaser}e).
    
  \item\textbf{Coils w/ amplitude (CLA)}: Instead of the number of
    coils, we use the amplitude of the coils with the interpretation
    that more time to compress means that the individual coils will be
    longer, akin to folding paper (Figure~\ref{fig:teaser}f).
  \end{itemize}	
\end{itemize}

Even if these visual representations appear reasonable and
appropriate, it is difficult to theoretically determine which of them
would be most effective for conveying the time scaling in a compressed
timeline.  For this reason, we conduct a user study to explore this
question empirically.

\subsection{Implementation Notes}

We implemented the EventLines technique using the D3 web-based
visualization toolkit~\cite{Bostock2011}, including all six of our
proposed timeline representations.  Our motivation for implementing
all six was to be able to rigorously evaluate all representations.
Our basic implementation is based on the D3 \texttt{axis} and
\texttt{scale} objects, allowing for easily including the EventLines
technique into an existing time-series visualization with minimal
changes to the current source code.  In total, our prototype toolkit
consists of 1,288 lines of JavaScript code, and it is similar to a D3
function call that requires passing the event data and setting the
width and height of the visualization.  There is a predefined maximum
height for a rectangle, curve or coils, and a predefined maximum
number for coils and dash which can be changed by the users.

While our EventLines implementation is currently only a research
prototype and is not production ready, we plan on releasing the
software as Open Source on the Internet as soon as possible.

%% file: contents/04-studies.tex

\section{Crowdsourced User Study}

We have reached a point in this work where we have a radical approach
to managing clutter in dense event timelines, but we do not know
whether the proposed visual representation to convey varying time
scaling will be powerful enough to support the new approach.  More
specifically, we do not know which of the six axis
representations---if any---will be able to convey the time axis
scaling to users of the EventLines time distortion technique.  To
begin to explore these questions, we conducted a crowdsourced user
study on graphical perception tasks in timeline visualizations.  In
the below section, we first present our pilot study that we conducted
to calibrate the evaluation.  We then describe the design rationale
for the study itself.  This is followed by the dataset generation, the
crowdsourced participants, task, and the procedure.

\subsection{Pilot Study}

In order to develop a general idea of how people receive our visual
representations and to determine which is the best visual for
estimating the amount of time compression, we conducted a pilot study
on Amazon Mechanical Turk using multiple choice tasks and open
questions.  One question asked users of their general impression of
our different visual representations.  We also included several tasks
requesting users to estimate the amount of time compression in a
visualization, using the same representation with different amounts of
time compression in three trials.

We recruited 10 participants for each visual representation after
removing those participants who worked on multiple representations to
eliminate the learning effect.  After calculating the correctness of
the results, we found the following high-level results:

\begin{enumerate}
\item Given the information that our visual representation is related
  to time, and asked what user thinks of them without any labels on
  the time axis, coil-like glyph representations achieved better
  awareness of time compression; and
\item Transparency of axis appearance performed worst for time
  compression estimation.
\item For other time comparison or estimation task, number coils
  (CLN), stipples (STP), and rectangles (RCT) have better performance
  than others.
\end{enumerate}

\subsection{Design Rationale}

Based on our goals and the results from our pilot study, we made
several design decisions for our study:

\begin{itemize}
\item\textbf{Task}: We decided to refine our study to one type of task
  with 10 trials so that we can cover as many combinations as possible
  and exclude deadwood Turkers~\cite{ElmqvistYi2012} according to the
  consistency of participants' answers.
  
\item \textbf{Crowdsourcing}: Our experience from running the pilot
  study as a crowdsourced graphical perception experiment was
  positive: we detected few deadwood Turkers, and results were
  consistent with our informal in-person pilots.
  
\item \textbf{No training}: We constrained each participant so that
  they can only finish one type of representation, avoiding learning
  effects between visual representations.
  
\end{itemize}

Our hypothesis was that some visual representations are superior to
others for comparing the amount of time compression.  To further
divide these representations into groups, we designed a task where
participants determined the interval that represented the most amount
of time compression among three labeled intervals in an axis
representation.

\subsection{Dataset Generation}


For our user study, we wanted the type of bursty discrete event data
that our visualization techniques were designed for.  To achieve this,
we generate datasets using a power law distribution according to this
theory and the inverse equation (1).  In the below equation, $x_0$ and
$x_1$ constrain the range of our generated data to the interval
($x_0$, $x_1$).  Our generated uniform distributed data is $y$ and $X$
is the desired output data.  This means for 100 uniform distributed
events $y$, we can generate 100 events $X$ using the power law
distribution.  Here, $n$ controls the distribution power.  That is,
for a small value of $n$, the number of data at each time value is
small, which distributes the data evenly, while a large value of $n$
will produce more bursty data.
After several trials with different parameter settings, we finalize
the total number of event data to 100 events, and $n$ to 37 which
generated the best distribution that meets our requirement.
\begin{equation}
X=[(x_1^{n+1} - x_0^{n+1}) * y + x_0^{n+1}]^{1/(n+1)}
\end{equation}

\subsection{Participants}

A total of 123 participants (62 female) completed our study on Amazon
Mechanical Turk.  We limited the demographic to the United States in
order to eliminate the influence of language.  To ensure the quality
of the workers, we required an approval rating of more than 0.95 and
more than 1,000 approved HITs.  83 people had more than bachelor's
degree, with 6 people having high school degrees.  We eliminated
random clickers by removing any trials where the completion time was
shorter than a reasonable time (2 seconds) for each task, and
eliminated duplicate participants through their worker ID, IP address,
and demographic information.  This yielded a total of 90 participants,
with 15 assigned to each visual representation.

\subsection{Experimental Design}

\begin{figure}[htb]
  \includegraphics[width=\linewidth]{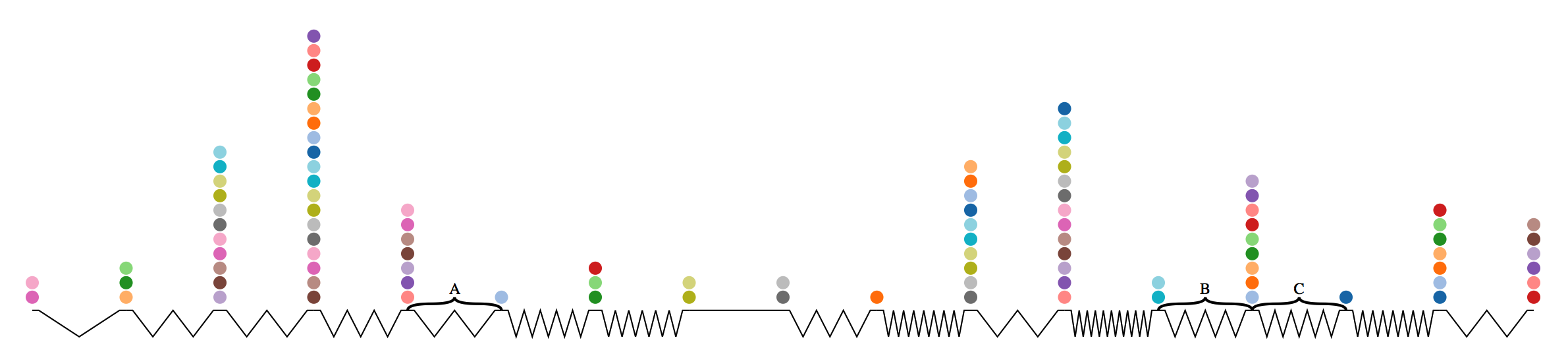}
  \caption{\textbf{User study trial.}
  A trial of our user study using the Coil /w numbers (CLN) with 100 data points.}
  \label{fig:solution}
\end{figure}

The goal of our study was to distinguish the efficiency of comparing
the amount of time compression between six visual representations.
Therefore, the types of visual representations are the most important
factor, and we thus divided the experiment into six blocks, one for
each representation.  Another important factor is the dataset and the
intervals selected for comparison.  To avoid the random effect of
data, each block had the same task, the same event data and selected
intervals, and the same number of trials organized in the same order.
We used our dataset generation method to generated 10 series of event
data and another random mechanism to generate three labeled intervals.
One trial of our Coils /w Numbers (CLN) experiments is shown in
Figure~\ref{fig:solution}.  In other words, the experiment uses the
one same task with 10 trials for each visual representation.  Some
answers are obvious, while others are hard to solve for some visual
representations.  Participants were restricted to only work on one
visual representation with 10 datasets.  The dependent variables we
measured are correctness and completion time for each trial because we
believe these two variables may accurately reflect the efficiency of
the visual representations for specific trials.

\subsection{Procedure}

We used the Qualtrics survey platform to create our experiments.
After generating each dataset, we made screenshots and created images
for each visual representation, and then used these images to create
questions for each trial.  We embedded our experiments as an embedded
frame on Amazon's Mechanical Turk platform.  After participants
accepted the HIT they viewed, participants were shown the consent form, followed
by a statement describing our payment policy after accepting.  Then,
an instruction page described our visual representation including what
the dots, the horizontal axis, and the visual representation mean, and
in what situation one interval represents more amount of time
compression than the other.  This was followed by a training block
consisting of four questions similar to the real tasks.  Participants
who did not complete the training questions correctly were not allowed
to continue the study.

After the four training questions, participants started the experiment
with an image showing one of our visual representation, a question
that asks ``Which interval represents more time?'', and three choices.
Each trial consisted of a single webpage and a timer to record the
completion time (not shown to participants).  Participants had to
answer each question to proceed, but could stop the experiment at any
time by closing the browser tab.  After ten trials of this question,
the task was finished and participants were asked to fill out a
demographic questionnaire.  At the end, participants were requested to
copy a code generated on Qualtrics and paste it into a textbox below
the embedded frame for MTurk payment.

%% file: contents/05-results.tex

\section{Results}

Our experiment used ten repetitions of one task for each visual
representation.  In the following treatment, we discuss correctness
and completion time for all visual representations.

\begin{figure*}[htb]
  \begin{minipage}{.32\linewidth}
    \includegraphics[width=0.95\textwidth]{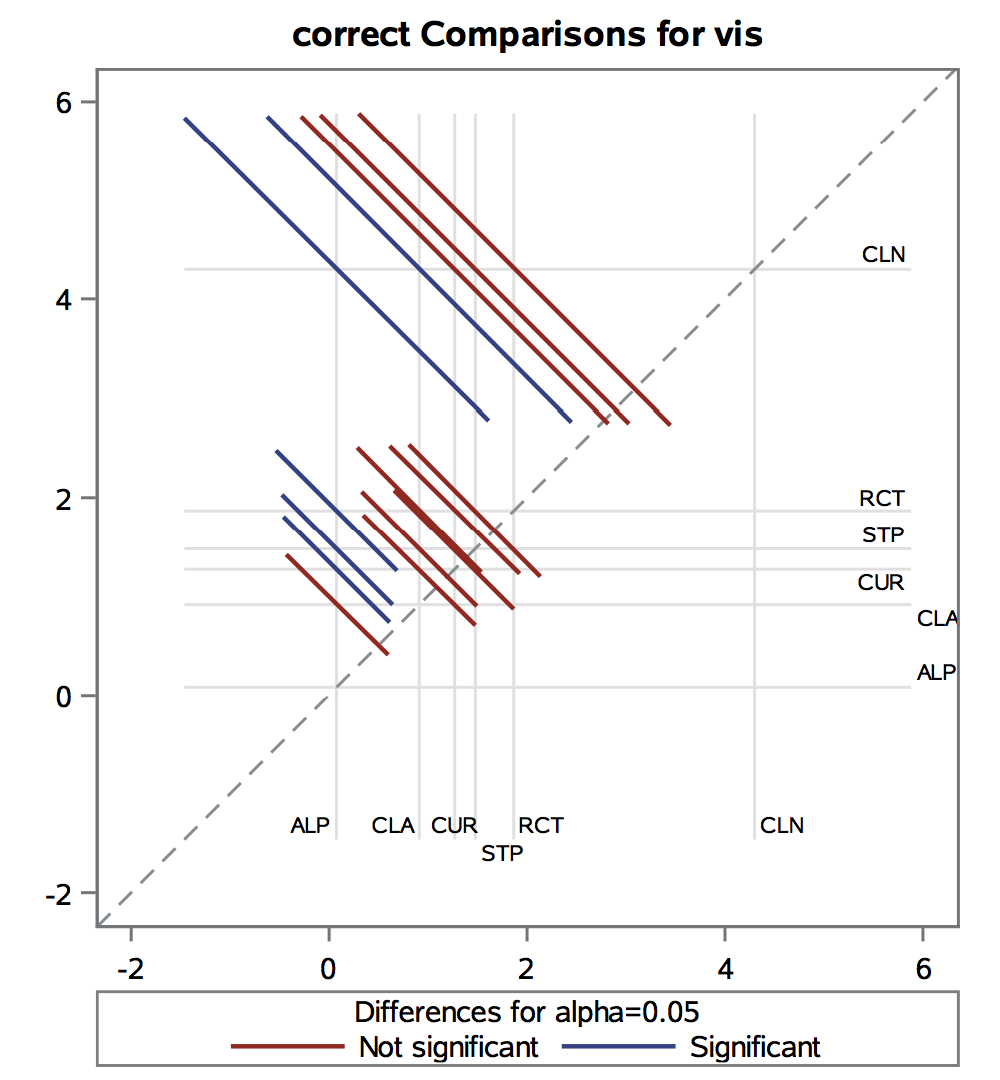}
    \captionof{figure}{Difference plot for 6 visual types (outliers
      removed).}
    \label{fig:results1}
  \end{minipage}\hfill%
  \begin{minipage}{.32\linewidth}
    \includegraphics[width=0.95\textwidth]{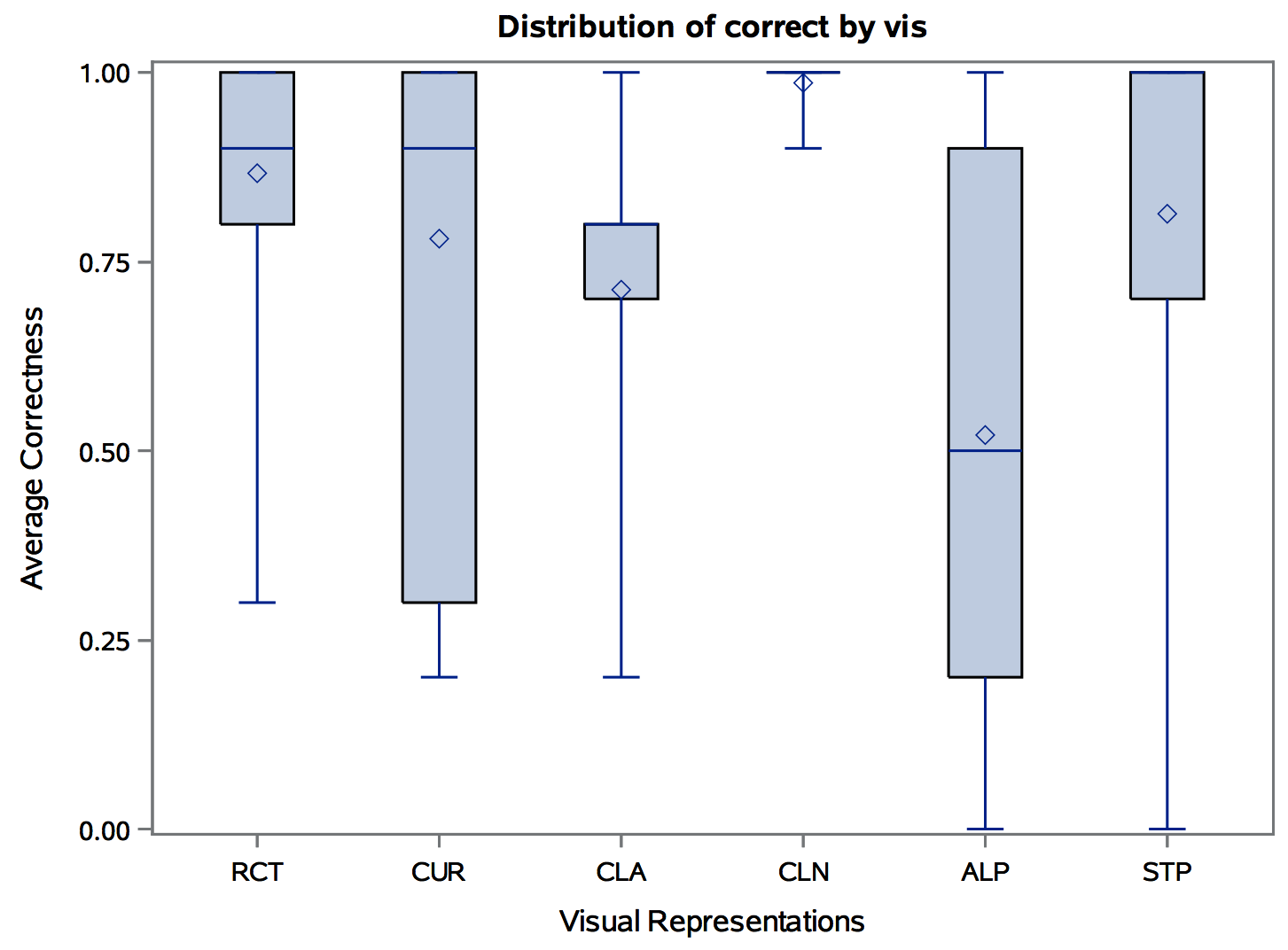}
    \captionof{figure}{Boxplot of average correctness vs.\ visual
      representations (outliers removed).}
    \label{fig:results2}
  \end{minipage}\hfill%
  \begin{minipage}{.32\linewidth}
    \includegraphics[width=0.95\textwidth]{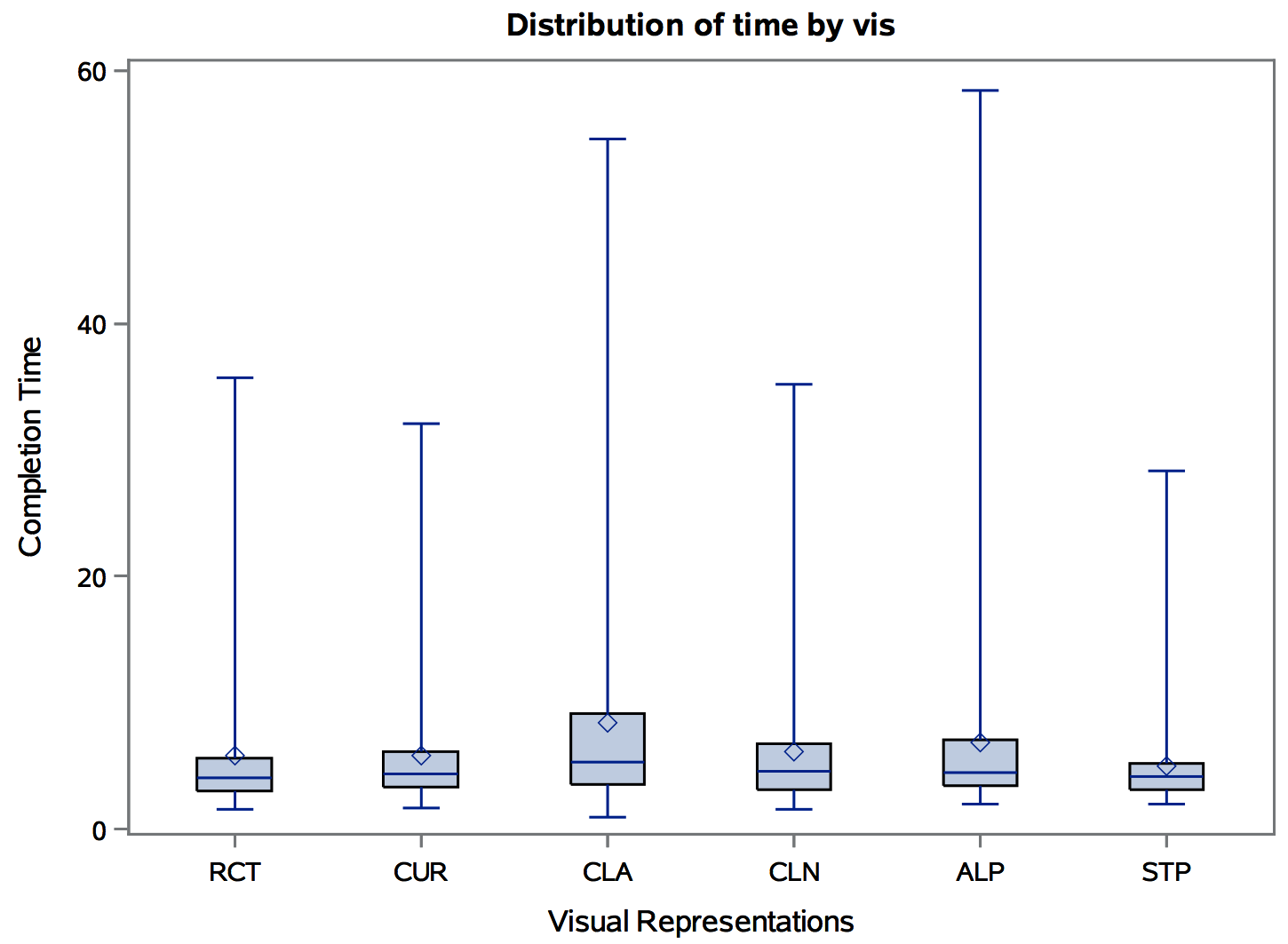}
    \captionof{figure}{Boxplot of completion time vs.\ visual
      representations (outliers removed).}
    \label{fig:results3}
  \end{minipage}
\end{figure*}

\paragraph{Correctness.}

We analyzed the main effect of Visual Representation Type ($V$) on correctness using logistic regression.
Figure~\ref{fig:results1} reveals the differences in correctness between pairs of visual representations.
Our analysis shows that the Transparency representation (ALP) differs significantly from all other visual representations except Coils with Amplitude (CLA).
Additionally, Coils with Numbers (CLN) demonstrates significant differences compared to other representations.
However, we found no statistical basis for clearly separating the six visual representations into distinct groups.
When analyzing the difficulty effect (easy versus hard trials), we found that hard trials showed no discernible patterns.
Further analysis of correctness versus visual representation ($V$) for hard trials alone revealed that Coils with Numbers (CLN) and Transparency (ALP) formed two distinct groups ($p < 0.05$), while other types showed no significant differences.
Figure~\ref{fig:results2} presents a boxplot illustrating the average correctness for each visual representation.

\paragraph{Completion Time.}
We measured trial completion time from the moment participants entered the trial page until their final click before submission.
Given time's continuous nature, we employed a parametric repeated-measures analysis of variance (RM-ANOVA) for our analysis.
The results indicate that Coils with Amplitude (CLA) differs significantly from all visual representations except Transparency (ALP).
Figure~\ref{fig:results3} displays a boxplot of completion times for each visual representation.

%% file: contents/06-discussion.tex
\section{Discussion}

In summary, we found that Coils /w Numbers (CLN) performs better than
Transparency (ALP) for low time compression, while there is no
distinction for all six visual representations if the difference of
compression is obvious.

\subsection{Explaining the Results}

Before running our experiment, we anticipated that the results would
allow us to divide the six visual representations into two groups:
Transparency (ALP) and Coils /w Amplitude (CLA) in one low-performing
group, as indicated by the pilot study, while the others would form
another group.  However, from our correctness analysis, we were only
able to distinguish Transparency (ALP) and Coils /w Numbers (CLN),
although there is significant difference between Amplitude (CLA) and
Coils /w Numbers (CLN).  The reason may be that the easy repetitions
did not provide any discrepancy effect on the correctness.

One surprising finding is there is a learning effect on time.
We analyzed the response time of the first two repetition data and
found that Coils /w Amplitude (CLA) was significantly different
compared to other representations.  We think that Coils /w Amplitude
(CLA) simply requires longer time for users to learn.

\subsection{Generalizing the Results}

Our study, while fairly comprehensive, only includes one of many
potential tasks that users may want to perform on discrete timelines:
comparing the amount of time compression.  If we compare our new
visual representations with traditional timeline visualization, the
latter will be trivially superior, simply by virtue of having no time
compression!

Although our visual representations are based on time event data, our
technique should be able to generalize to any bursty data visualized
using line chart visualization.  However, a data-aware visual
representation does not always succeed.  If the data is uniformly
sensed and not irregular, our representations will perform similar or
worse compared to traditional timeline visualization.  However,
irregular data is a common thing in our daily lives, as indicated by
the examples given by~\cite{Newman2005}.

Our user study explicitly did not compare time compression techniques
to standard linear timelines, and this was a deliberate decision on
our behalf.  It is important to note that our EventLines technique is
not a replacement to normal linear timelines.  Instead, it is a
complement that can be used in certain situations for bursty event
data.  In fact, a conceivable use of EventLines may be to trigger it
on demand.  In other words, a timeline visualization may have a toggle
that switches back and forth between a regular linear timeline, and an
EventLines compressed timeline with a temporal axis visual
representation to convey this fact.

%% file: contents/07-application.tex

\section{Application: Search in the DIA2 Platform}

We integrated our Coils /w Numbers (CLN) visual representation along with a traditional timeline visualization into a search widget on the
DIA2 (Deep Insights Anytime, Anywhere) platform, a web-based visual analytics system for program managers and staff at the U.S.\ National Science Foundation~\cite{Madhavan2014}.

\begin{figure}[htb]
  \centering
  \includegraphics[width=\linewidth]{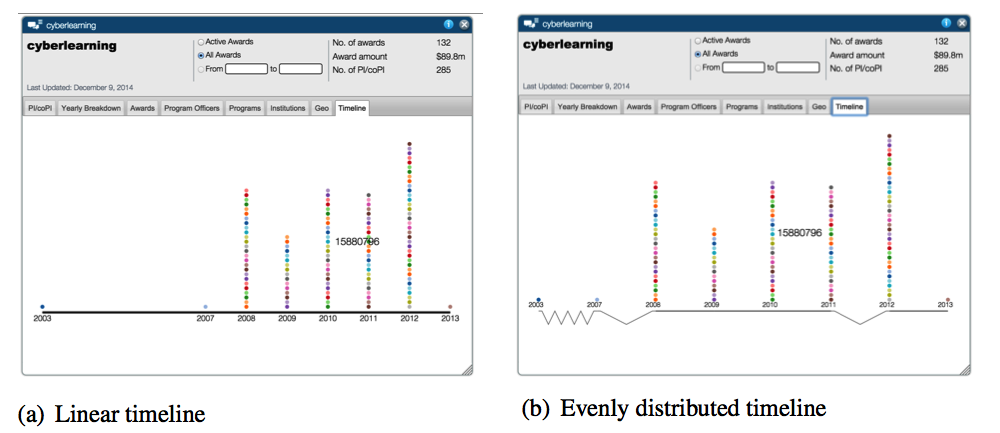}
  \caption{\textbf{EventLines in the DIA2 system.}
  (a) Search result timeline; (b) EventLines integrated with the timeline.}
  \label{fig:dia2}
\end{figure}

The search feature in DIA2 can return a timeline of results for specific events.
Figure~\ref{fig:dia2} shows how EventLines has been integrated with the search result timeline, evenly distributing the events in space.
This eases selection and allows tooltip information about each event to be clearly seen.

%% file: contents/08-conclusion.tex

\section{Conclusion \& Future Work}

We have presented EventLines, a data-aware time distortion technique
for visualization discrete event datasets where the axis scaling is
automatically adapted to maximize screen space usage.  This will yield
a more uniform visual complexity across the timeline visualization and
will accordingly make the event data easier to view and understand.
However, this time distortion comes at a cost: the time axis no longer
has a global linear scale.  To address this problem, we propose six
different visual representations of the time axis that convey the
varying scaling based on appearance, glyphs, and thickness.  A
crowdsourced user study found that of the six techniques, timelines
using coils with different numbers are superior.

Our application example for search in the DIA2 platform shows some of the 
potential for the EventLines method, but we envision applying the idea to a 
much wider set of problems, both inside as well as outside the DIA2 platform. 
For example, we are planning to use the technique to visualize the discrete events
that take place during the history of a funded research award, such as
publications, change requests, PI transfer, etc.  Beyond this, we
anticipate using the idea to visualize the timeline of discrete events
for people's careers, such as publications, grants funded, students
graduated, awards received, and ranks attained for faculty members (in
fact, this was the motivating scenario for this work in the first
place).